\documentclass[12pt]{article}
\begin{document}
\begin{titlepage}

\title{Distorted Torsion Tensor, Teleparallelism and Spin 2 Field Equations}

\author{J. W. Maluf$\,^{*}$ \\
Instituto de F\'{\i}sica, 
Universidade de Bras\'{\i}lia\\
70.919-970 Bras\'{\i}lia DF, Brazil\\}
\maketitle
\bigskip
\bigskip

\begin{abstract}
A notion of distorted torsion tensor was introduced by Okubo, in the 
establishment of the Nijenhuis-Bianchi identity and of BRST-like operators.
These quantities are constructed with the help of the Nijenhuis tensor, 
which in turn is defined in terms of a (1,1) tensor $S^\lambda_\mu$. This 
tensor enters the construction of the distorted torsion tensor. We use this
tensor to extend the teleparallel equivalent of general relativity (TEGR) 
into a theory defined by the tetrad fields and by the 
tensor $S^\lambda_\mu$. The ordinary TEGR is recovered if 
$S^\lambda_\mu=\delta^\lambda_\mu$. We consider the flat space-time 
formulation of the theory, in terms of $S^\lambda_\mu$ only, and show that
this tensor satisfies the wave equation for massless spin 2 fields.
\end{abstract}
\thispagestyle{empty}
\bigskip
\vfill
\begin{footnotesize}
\end{footnotesize}

\bigskip
{\footnotesize
\noindent {*} wadih@unb.br, jwmaluf@gmail.com}
\end{titlepage}

\newpage
\section{Introduction}
The tools of differential geometry have been applied to the study of 
dynamical integrable systems \cite{Okubo1,Okubo2}. In this context, a quantity
that plays a major role is the Nijenhuis tensor \cite{NJ,Nakahara}, which  
has been considered in connection to the formulation of integrable 
models \cite{Das}. The vanishing of the Nijenhuis tensor yields interesting 
properties in manifolds with dual symplectic structures. One of these 
properties is the emergence of a number of conserved quantities in
involution, that are necessary for the complete integrability of certain
dynamical systems \cite{Okubo1,Okubo2}. The Nijenhuis tensor is defined in any
differentiable manifold $M$ of dimension D. In components, it reads
\cite{Nakahara}

\begin{equation}
N^\lambda_{\mu\nu}=S^\alpha_\mu \partial_\alpha S^\lambda_\nu-
S^\alpha_\nu \partial_\alpha S^\lambda_\mu -
S^\lambda_\alpha(\partial_\mu S^\alpha_\nu-\partial_\nu S^\alpha_\mu)\,,
\label{1}
\end{equation}
where $S^\lambda_\mu$ is a mixed (1,1) tensor on $M$ that is usually related
to the dual symplectic structures on the manifold. 
The tensor $N^\lambda_{\mu\nu}$ is  entirely independent of any connection. 
In the context of
complex manifolds, the vanishing of the Nijenhuis tensor is related to the 
existence of integrable almost complex structures \cite{Nakahara}. We are not
aware of the applicability of this tensor in the framework of gravity 
theories.

One motivation for considering the Nijenhuis tensor in a geometrical framework
is the construction of BRST-like operators. BRST operators \cite{BRST} are the
building blocks of the called BRST quantization, which is specially applied to
gauge theories. In particular, this quantization approach yields a rigorous 
canonical quantization of Yang-Mills theories. Okubo \cite{Okubo3,Okubo4} 
investigated the existence of BRST-like operators in certain manifolds. He 
found that the existence of these operators depends (i) on the vanishing of 
both the Nijenhuis and Riemann tensors, and (ii) on a geometrical identity
which he calls the Nijenhuis-Bianchi identity. The construction of this
identity is based on an extension of the standard torsion tensor. Okubo 
defines distorted torsion tensors of the first and second kinds. Here, we will
consider the distorted torsion tensor of the first kind only. It is 
established with the help of the $S^\lambda_\mu$ tensor, and reads
\cite{Okubo3}

\begin{eqnarray}
{\cal T}^{\lambda}_{\mu\nu}&=&\partial_\mu S^\lambda_\nu-
\partial_\nu S^\lambda_\mu +\Gamma^\lambda_{\mu\sigma}S^\sigma_\nu-
\Gamma^\lambda_{\nu\sigma}S^\sigma_\mu \nonumber \\
&=& \nabla_\mu S^\lambda_\nu-\nabla_\nu S^\lambda_\mu+
S^\lambda_\sigma T^\sigma_{\mu\nu}\,,
\label{2}
\end{eqnarray}
where $T^\sigma_{\mu\nu}=\Gamma^\sigma_{\mu\nu}-\Gamma^\sigma_{\nu\mu}$, and
$\Gamma^\sigma_{\mu\nu}=e^{a\sigma}\partial_\mu e_{a\nu}$ 
is the Weitzenb\"ock connection. The covariant derivative 
$\nabla_\mu$ is constructed out of $\Gamma^\sigma_{\mu\nu}$.
Note that ${\cal T^\lambda_{\mu\nu}}$ reduces to $T^\lambda_{\mu\nu}$ if
$S^\lambda_\mu =\delta^\lambda_\mu$. \par 
\bigskip
\noindent Notation: $\mu, \nu, \,...$ are space-time indices and run from $0$
to $3$, $\mu=(0,i)$, where $i=1, 2, 3$; $a, b, c, \,...$ are Lorentz (tangent
space) indices and also run from $0$ to $3$, $a=((0),(i))$.\par
\bigskip
In this article we will investigate the consequences of the distorted torsion 
tensor (\ref{2}) in the framework of the teleparallel equivalent of general 
relativity (TEGR). For this purpose, we will abandon the usual relation of the
tensor $S^\lambda_\mu$ with the dual symplectic structures of integrable 
models (if $f_{\mu\nu}=-f_{\nu\mu}$ and $F_{\mu\nu}=-F_{\nu\mu}$ are two
symplectic forms on a manifold, then one may identify 
$S^\lambda_\mu=f_{\mu\rho}F^{\rho\lambda}$ in the context of integrable 
dynamical systems \cite{Okubo1}). We will treat $S^\lambda_\mu$ as a regular
$(1,1)$ tensor. Quantities similar to 
eq. (\ref{2}) may possibly generalize the TEGR to theories that incorporate
spin 1 or spin 2 fields with non-trivial couplings with the tetrad fields. 

In section 2 we establish the first Lagrangian formulation of the extended 
TEGR, based on the distorted torsion tensor (\ref{2}), and obtain the field 
equations for both the tetrad fields and the tensor $S^\lambda_\mu$. 
In order to probe the nature of the tensor $S^\lambda_\mu$ in the present 
gravitational context,
we consider a TEGR-type theory with only the $S^\lambda_\mu$
tensor, i.e., in flat space-time. It turns out that this tensor obeys the
standard wave equation, without assuming any linearisation of the field 
variables. Thus, in this simplified formulation, the tensor  $S^\lambda_\mu$
describes propagating massless spin 2 fields. We discuss
some features and limitations of this model, regarding the flat space-time
limit, and then in sections 3 and 4 we present the second and third models
where the flat space-time limit is obtained in a more conventional way. 
Finally, in section 5 we present the final comments.

\section{The extended TEGR - the first model}

Teleparallelism is a geometrical framework where the notion of distant
parallelism is well defined. 
In order to understand this notion, let us consider
a vector field $V^\mu(x)$ in space-time. The projection of this vector on a
certain frame, in the tangent space at the position $x^\alpha$, is given by 
$V^a(x^\alpha)=e^a\,_\mu(x^\alpha)\,V^\mu(x^\alpha)$. At the position
$x^\alpha + dx^\alpha$, we have $V^a(x^\alpha+dx^\alpha)=
e^a\,_\mu(x^\alpha+dx^\alpha)\,V^\mu(x^\alpha+dx^\alpha)$. It is easy to show
that $V^a(x^\alpha)$ and $V^a(x^\alpha+dx^\alpha)$ are parallel, i.e.,
$V^a(x^\alpha)=V^a(x^\alpha+dx^\alpha)$, if the covariant derivative
$\nabla_\mu V^{\lambda}$ vanishes. This covariant derivative is constructed 
out of the Weitzenb\"ock connection. Therefore, the equation 
$\nabla_\mu V^{\lambda}=0$ establishes a condition of distant parallelism in
space-time, and the Weitzenb\"ock connection plays a crucial role in this
concept. It is easy to verify that the tetrad fields are auto-parallel,
i.e., $\nabla_\mu e_a\,^\lambda\equiv 0$.

The ordinary formulation of the TEGR is obtained by means of a geometrical 
identity between the scalar curvature $R(e)$ constructed out of a set of 
tetrad fields $e^a\,_\mu$, and a quadratic combination of the torsion tensor 
of the Weitzenb\"ock connection, 
$T_{a\mu\nu}=\partial_\mu e_{a\nu}-\partial_\nu e_{a\mu}
\equiv e_a\,^\lambda T_{\lambda \mu\nu}$,

\begin{equation}
eR(e) \equiv -e\left({1\over 4}T^{abc}T_{abc} + 
{1\over 2}T^{abc}T_{bac} - T^{a}T_{a}\right)
+ 2\partial_{\mu}(eT^{\mu})\,,
\label{3}
\end{equation}
where $ T_{a} = T^{b}\,_{ba}$, 
$T_{abc} = e_{b}\,^{\mu}e_{c}\,^{\nu}T_{a\mu\nu}$, and 
$e=\det(e^a\,_\mu)$. Neglecting the total
divergence, the Lagrangian density for the TEGR in asymptotically flat 
space-times is given by \cite{Maluf1,Maluf2}

\begin{eqnarray}
L(e) &=& -k\,e\left({1\over 4}T^{abc}T_{abc} + {1\over 2}T^{abc}T_{bac} 
- T^{a}T_{a}\right) - {1\over c}L_{M}\nonumber \\
& \equiv & -ke\Sigma^{abc}T_{abc} - {1\over c}L_{M}\,,
\label{4}
\end{eqnarray}
where $k = c^3/(16\pi G)$, $L_{M}$ represents the Lagrangian density for the 
matter fields, and $\Sigma^{abc}$ is defined by \cite{Maluf1}
\begin{equation}
\Sigma^{abc} = {1\over 4}\left(T^{abc} + T^{bac} - T^{cab}\right) 
+ {1\over 2}\left(\eta^{ac}T^{b} - \eta^{ab}T^{c}\right)\,.
\label{5}
\end{equation}

The extended, generalized form of the TEGR is constructed by replacing 
$T_{\lambda\mu\nu}$ with ${\cal T}_{\lambda\mu\nu}$, the latter given by 
eq. ({\ref{2}). Thus, the vacuum formulation of the theory is determined by
the Lagrangian density constructed out of $e_{a\mu}$ and of the tensor
$S^\lambda_\mu$, and reads

\begin{equation}
L(e_{a\mu},S^\lambda_\mu)=-k\Sigma^{abc}{\cal T}_{abc}\,,
\label{6}
\end{equation}
where $\Sigma_{abc}$ is now given by
\begin{equation}
\Sigma^{abc} = {1\over 4}\left({\cal T}^{abc} + {\cal T}^{bac} 
- {\cal T}^{cab}\right) 
+ {1\over 2}\left(\eta^{ac}{\cal T}^{b} - \eta^{ab}{\cal T}^{c}\right)\,.
\label{7}
\end{equation}

The variations of the Lagrangian density (\ref{4}) with respect to $e^{a\mu}$
and $S^\lambda_\mu$ yield, respectively, 

\begin{eqnarray}
&{}&e_{b\mu}\partial_\sigma(e\Sigma^{b\sigma\nu}S^\rho_\nu\,e_{a\rho})
-{1\over 2} (e\Sigma_\mu\,^{\sigma\nu}
e_{a\lambda}{\cal T}^\lambda\,_{\sigma\nu}) \nonumber \\
&{}&+e\Sigma_{\lambda a}\,^\nu {\cal T}^\lambda\,_{\mu\nu} -
{1\over 4}e e_{a\mu} \Sigma^{bcd} {\cal T}_{bcd} =0\,,
\label{8}
\end{eqnarray}

\begin{equation}
\nabla_\sigma (e\Sigma_\lambda\,^{\sigma\mu})=
\partial_\sigma(e\Sigma_\lambda\,^{\sigma\mu})-
\Gamma^\rho_{\sigma\lambda}(e\Sigma_\rho\,^{\sigma\mu})=0\,.
\label{9}
\end{equation}

We note that if we enforce $S^\lambda_\mu=\delta^\lambda_\mu$ in 
eq. (\ref{8}), we obtain 

\begin{equation}
e_{a\nu}e_{b\mu}\partial_\sigma(e\Sigma^{b\nu\sigma})-
e(\Sigma_{\lambda a}\,^\nu T^\lambda\,_{\mu\nu}-
{1\over 4}e_{a\mu}\Sigma^{bcd}T_{bcd})=0\,,
\label{10}
\end{equation}
which are the vacuum space-time field equations for the tetrad field in the
TEGR, and which are equivalent to Einstein's equations in vacuum 
\cite{Maluf2}. However, this enforcement ($S^\lambda_\mu=\delta^\lambda_\mu$)
is not so trivial, as we will see below.

In order to probe the nature of the field $S^\lambda_\mu$ in the equations 
above, let us assume $e^a\,_\mu(t,x,y,z)=\delta^a_\mu$ in the field equation
(\ref{9}). We remark that this is not a flat space-time {\it limit} of the
theory, but just a flat space-time formulation of a TEGR-type theory, where
eq. (\ref{2}) reduces to
${\cal T}^{\lambda}_{\mu\nu}=\partial_\mu S^\lambda_\nu-
\partial_\nu S^\lambda_\mu$, and eqs. (\ref{6}) and (\ref{7}) remain 
unchanged. Thus, this is a {\it new theory} defined in flat space-time.
In this theory, the field equations for $S^\lambda_\nu$ are

\begin{equation}
\nabla_\rho (e\Sigma_\lambda\,^{\rho\nu})=
\partial_\rho(\Sigma_\lambda\,^{\rho\nu})=0\,.
\label{11}
\end{equation}
Before we proceed, let us establish the following convention:

\begin{equation}
S^\lambda_\mu \longrightarrow S^\lambda\,_\mu\,.
\label{12}
\end{equation}
As a consequence, we have $S^{\lambda\mu}=\eta^{\mu\rho} S^\lambda\,_\rho$,
and $S_{\alpha \mu} =\eta_{\alpha \lambda} S^\lambda\,_\mu$, where 
$\eta_{\mu\nu}=(-1,+1,+1,+1)$ is the flat Minkowski metric tensor. With
the help of the notation above, and in view of eq. (\ref{7}), 
the field equations (\ref{11}) read

\begin{eqnarray}
\partial_\rho \Sigma^{\lambda \rho \nu}&=&
{1\over 4} (\partial_\rho \partial^\rho) (S^{\nu\lambda} 
+S^{\lambda\nu} ) \nonumber \\
&{-}&{1\over 4}(\partial_\rho \partial^\nu S^{\lambda\rho}
+\partial_\rho \partial^\lambda S^{\nu\rho}
+\partial_\rho \partial^\lambda S^{\rho\nu}
+\partial_\rho \partial^\nu S^{\rho\lambda} ) \nonumber \\
&+&{1\over 2} (\eta^{\lambda\nu}\partial_\rho\partial_\sigma S^{\rho\sigma}
-\eta^{\lambda\nu}\,(\partial_\rho\partial^\rho) \,S^\sigma\,_\sigma
+\partial^\lambda\partial^\nu \,S^\rho\,_\rho) \nonumber \\
&=& 0\,,
\label{13}
\end{eqnarray}
where $\partial^\lambda =\eta^{\lambda\mu}\partial_\mu$.

We note that the left hand side of the equation above is naturally 
symmetric in the indices $(\lambda \nu)$. Therefore, the antisymmetric part
of the tensor $S^{\lambda\nu}$ is completely undetermined from the field
equations. For the time being, and without loss of generality, we will assume
in the present context that the tensor $S^{\lambda\nu}$ is symmetric, i.e.,
$S^{\lambda\nu}=S^{\nu\lambda}$.

Equation (\ref{13}) may be simplified by first implementing the symmetry
condition $S^{\lambda\nu}=S^{\nu\lambda}$, and then by contracting it with
the metric tensor $\eta_{\lambda\nu}$. This last operation yields

\begin{equation}
(\partial_\lambda \partial^\lambda) \,S^\rho\,_\rho=
\partial_\rho\partial_\lambda S^{\rho\lambda}\,.
\label{14}
\end{equation}
As a consequence, eq. (\ref{13}) simplifies to

\begin{equation}
(\partial_\rho\partial^\rho)\,S^{\lambda\nu}
-\partial_\rho\partial^\nu S^{\lambda\rho}
-\partial_\rho\partial^\lambda S^{\nu\rho}
+\partial^\lambda \partial^\nu S^\rho\,_\rho=0\,.
\label{15}
\end{equation}
It is interesting to note that the left hand side of the equation above is
similar to the linearised Ricci tensor for $h_{\mu\nu}$, which is considered
in the investigation of linearised gravitational waves, where
$g_{\mu\nu} \simeq\eta_{\mu\nu}+h_{\mu\nu}$ (see eq. (20.10) of 
Ref. \cite{Ray}). In contrast, eq. (\ref{15}) is exact, as it does not follow
from any linearisation procedure.

It is very easy to verify that eq. (\ref{15}) is invariant under the gauge
transformation 

\begin{equation}
{S^\prime}^{\lambda\mu}=S^{\lambda\mu}+\partial^\lambda V^\mu
+\partial^\mu V^\lambda\,,
\label{16}
\end{equation}
where $V^\lambda(x)$ is an arbitrary vector field. This vector field may be 
used to fix the de Donder, or Einstein, or Hilbert, or Fock gauge \cite{Ray},

\begin{equation}
\partial_\rho {S^\prime}^{\lambda\rho}-
{1\over 2}\partial^\lambda{S^\prime}^\rho\,_\rho=0\,,
\label{17}
\end{equation}
which is obtained by requiring the vector field $V^\lambda$ to satisfy

\begin{equation}
(\partial_\rho\partial^\rho)\,V^\lambda=
-\partial_\rho S^{\lambda\rho}+{1\over 2} \partial^\lambda S^\rho\,_\rho\,.
\label{18}
\end{equation}
Formally, there exists a solution to the equation above for $V^\lambda$,
and therefore the gauge (\ref{17}) is consistent. With the imposition of this
gauge, the field equations (\ref{15}) reduce to 

\begin{equation}
(\partial_\rho\partial^\rho) {S}^{\lambda\nu}=0\,.
\label{19}
\end{equation}
where we have dropped the prime. This is the 
wave equation for a massless spin 2 field, but not for
a pure spin 2 field \cite{Watanabe}, as the trace $S^\rho\,_\rho$ is 
non-vanishing. Since $S^\lambda\,_\mu=\delta^\lambda_\mu$ is a
condition that leads eq. (\ref{8}) to eq. (\ref{10}), solutions of 
eq. (\ref{19}) may have the form

\begin{equation}
S_{\lambda \mu} = \eta_{\lambda \mu} +\texttt{wave solution}\,.
\label{20}
\end{equation}

The formalism presented above allows to make a clear distinction between the
propagation of non-linear gravitational waves, that arise from 
eq. (\ref{10}), and of spin 2 fields (gravitons, in a quantum theory),
an issue that is not 
satisfactorily addressed in the standard formulation of
general relativity. In the framework of the coupled field equations
(\ref{8}) and (\ref{9}), a source of the field $S^{\lambda\mu}$ is the
dynamical geometry of the space-time, represented by time dependent
tetrad fields $e^a\,_\mu$. 

It is certainly not easy to obtain a solution for $S^\lambda\,_\mu$ of the 
coupled equations (\ref{8}) and (\ref{9}). It is necessary to make some
simplifications and approximations to handle these equations. Equation  
(\ref{9}) is a wave equation for $S^{\lambda\,\mu}$ in the presence of a 
non-trivial set of tetrad fields, i.e., in a non-flat space-time. 
In the context
of the coupled equations (\ref{8}) and (\ref{9}), $S^\lambda\,_\mu$ should be
part of the geometry, as we conclude from eq. (\ref{20}). In particular, the
tensor

\begin{equation}
S_{ab}=e_a\,^\lambda e_b\,^\mu S_{\lambda \mu}\,,
\label{21}
\end{equation}
could be an effective metric tensor for the tangent space, i.e., an 
extension of the flat Minkowski (tangent space) metric tensor 
$\eta_{ab}=(-1,+1,+1,+1)$, which includes oscillations (fluctuations) of the
background geometry. This interpretation is physically possible at least in
the context of weak gravitational fields. Such background oscillations could
be similar to (continuous) gravitational waves that permeate the 
universe, and even to the background noises that are significant
in the detection of gravitational waves, but otherwise very weak to be 
characterized in ordinary circumstances. 

We note that 
$e^a\,_\mu=\delta^a_\mu$ and $S^\lambda\,_\mu=\delta^\lambda_\mu$ are,
together, solutions of the field equations (\ref{8}) and (\ref{9}). However,
it seems that neither $e^a\,_\mu=\delta^a_\mu$ alone, nor 
$S^\lambda\,_\mu=\delta^\lambda_\mu$ alone, are solutions of the field
equations. Thus, the flat space-time (the vacuum) is also determined by the 
condition $S^\lambda\,_\mu=\delta^\lambda_\mu$. By enforcing 
$S^\lambda\,_\mu=\delta^\lambda_\mu$, without requiring  
$e^a\,_\mu=\delta^a_\mu$ , eq. (\ref{9}) 
imposes further, additional restrictions on the tetrad fields, that may lead
to inconsistencies. 
It is not impossible that the framework 
described above is physically correct, but certainly it is not 
sufficiently convincing, at least at present. Thus, in the next section we
will address the issue of how to obtain the limit 
$S^\lambda\,_\mu\rightarrow\delta^\lambda_\mu$, independently of the 
flat space-time limit, by formulating a second, more conventional model.

\section{The second model}

The aim of this section is to speculate on another extended version of the 
TEGR where the limit $S^\lambda\,_\mu \rightarrow \delta^\lambda_\mu$ may
be obtained independently of the flat space-time limit 
$e^a\,_\mu \rightarrow \delta^a_\mu$. The converse does not hold, since
a non-vanishing spin 2 field, or any other form of matter fields, is 
always a source to the gravitational field equations. The model to be
discussed below is a more conservative description of the physical system.

Let us first establish the notation. In this section, we will use the
following definitions:

\begin{eqnarray}
\Sigma^{abc} &=& {1\over 4}\left(T^{abc} + T^{bac} - T^{cab}\right) 
+ {1\over 2}\left(\eta^{ac}T^{b} - \eta^{ab}T^{c}\right)\,,
\label{22}  \\
\Phi^{abc} &=& {1\over 4}\left({\cal T}^{abc} + {\cal T}^{bac} 
- {\cal T}^{cab}\right) 
+ {1\over 2}\left(\eta^{ac}{\cal T}^{b} - \eta^{ab}{\cal T}^{c}\right)\,, \\
\label{23}
T^a\,_{\mu\nu}&=&\partial_\mu e^a\,_\nu-\partial_\nu e^a\,_\mu\,, 
\label{24} \\
{\cal T}^{\lambda}\,_{\mu\nu}&=&\nabla_\mu S^\lambda\,_\nu-
\nabla_\nu S^\lambda\,_\mu \,.
\label{25}
\end{eqnarray}
Equation (\ref{25}) replaces eq. (\ref{2}). As before, the tetrad 
fields convert space-time into Lorentz indices, and vice-versa.
The covariant derivatives are constructed out of the Weitzenb\"ock 
connection, $\Gamma^\lambda_{\mu\nu}=e^{a\lambda}\partial_\mu e_{a\nu}$.
Thus, we have

\begin{equation}
\nabla_\mu S^\lambda\,_\nu=\partial_\mu S^\lambda\,_\nu
+\Gamma^\lambda_{\mu\sigma} S^\sigma\,_\nu
-\Gamma^\sigma_{\mu\nu} S^\lambda\,_\sigma\,.
\label{26}
\end{equation}
Note that if we make $S^\lambda\,_\mu=\delta^\lambda_\mu$, the covariant
derivative vanishes. This feature is important for ensuring that in the limit
$S^\lambda\,_\mu \rightarrow\delta^\lambda_\mu$ we obtain the standard 
TEGR in vacuum, in the equations below.

The theory to be considered in this section is defined by the Lagrangian
density

\begin{equation}
L=-ke\Sigma^{abc}\,T_{abc}-ke\Phi^{abc}\,{\cal T}_{abc}\,.
\label{27}
\end{equation}
Variations of $L$ with respect to $S^\lambda\,_\mu$ and $e^{a\mu}$ yield
the field equations

\begin{equation}
\nabla_\mu(e\Phi_\lambda\,^{\mu\nu})=0\,,
\label{28}
\end{equation}
where

\begin{equation}
\nabla_\mu(e\Phi_\lambda\,^{\mu\nu})=\partial_\mu(e\Phi_\lambda\,^{\mu\nu})
-e\Gamma^\sigma_{\mu\lambda}\Phi_\sigma\,^{\mu\nu}
+\Gamma^\nu_{\mu\sigma}\Phi_\lambda\,^{\mu\sigma}\,,
\label{29}
\end{equation}
and

\begin{eqnarray}
&{}& e_{a\nu} e_{b\mu}\partial_\sigma(e\Sigma^{b\nu\sigma})-
e(\Sigma_{\lambda a}\,^\nu T^\lambda\,_{\mu\nu}-
{1\over 4}e_{a\mu}\Sigma^{bcd}T_{bcd}) \nonumber \\
&{=}&-{1\over 2}
e\,e_{a\lambda}\Phi_\mu\,^{\rho\nu}{\cal T}^\lambda\,_{\rho\nu}
+e\,\Phi^{\lambda\nu}\,_a\,{\cal T}_{\lambda\mu\nu}
-{1\over 4}e\,e_{a\mu}\Phi^{bcd}{\cal T}_{bcd} \nonumber \\
&{}&+\nabla_\rho\lbrack e\,e_{a\sigma}(\Phi_\mu\,^{\rho\nu} 
S^\sigma\,_\nu- \Phi_\nu\,^{\rho\sigma} S^\nu\,_\mu) \rbrack\,.
\label{30}
\end{eqnarray}
where

\begin{eqnarray}
\nabla_\rho\lbrack e(\Phi_\mu\,^{\rho\nu}S^\sigma\,_\nu-
\Phi_\nu\,^{\rho\sigma} S^\nu\,_\mu)e_{a\sigma}\rbrack&=&
\partial_\rho\lbrack e(\Phi_\mu\,^{\rho\nu}S^\sigma\,_\nu-
\Phi_\nu\,^{\rho\sigma} S^\nu\,_\mu)e_{a\sigma}\rbrack \nonumber \\
&-& \Gamma^\lambda_{\rho\mu}
\lbrack e(\Phi_\lambda\,^{\rho\nu}S^\sigma\,_\nu-
\Phi_\nu\,^{\rho\sigma} S^\nu\,_\lambda)e_{a\sigma}\rbrack\,.
\label{31}
\end{eqnarray}
The left hand side of the eq . (\ref{30}) is equivalent to Einstein's 
equations \cite{Maluf2}. In view of eq. (\ref{26}), we see that the right
hand side of this equation vanishes in the limit 
$S^\lambda\,_\mu \rightarrow\delta^\lambda_\mu$,
and thus we obtain the expected equations for the tetrad field in vacuum.
According to eq. (\ref{30}), the spin 2 field is a source to the 
gravitational field, as any other kind of matter fields.

By enforcing $e^a\,_\mu\rightarrow\delta^a_\mu$ in eq. (\ref{28}) (i.e., 
assuming that the theory is formulated in flat space-time in terms of
$S^\lambda\,_\mu$ only, as in section 2), we obtain from (\ref{28}) the 
equation

\begin{equation}
\partial_\mu\Phi_\lambda\,^{\mu\nu}=0\,,
\label{32}
\end{equation}
which reminds Maxwell's equations in covariant form, and
which is strictly equivalent to eq. (\ref{11}). Therefore, in the present
context, we also obtain eqs. (\ref{13}), (\ref{15}), and ultimately  eq. 
(\ref{19}),

\begin{equation}
(\partial_\rho\partial^\rho) {S}^{\lambda\nu}
=\biggl(\nabla^2-{1\over c^2}{{\partial^2}\over{\partial t^2 }}\biggr)
{S}^{\lambda\nu} =0\,.
\label{33}
\end{equation}
Thus, the field $S^{\lambda\mu}$ again satisfies the wave equation in the 
flat space-time formulation. In addition, the left hand side of 
eq. (\ref{28}) vanishes  
in the limit $S^\lambda\,_\mu \rightarrow\delta^\lambda_\mu$. This is the 
feature that characterizes the theory defined by  eq. (\ref{27}).

By contracting eq. (\ref{30}) with $e^a\,_\beta$, the left hand side of the 
equation becomes symmetric in the indices $\beta\mu$. This fact can be 
verified by transforming the resulting tensor into Einstein's tensor, which
is symmetric. However, it is not immediately clear that the right hand side
of (\ref{30}) is also symmetric. If it turns out that the right hand side of 
(\ref{30}) is not symmetric, the anti-symmetric part imposes conditions on
6 of the 16 components of the tetrad fields. This issue might be a novel 
feature, and will be investigated elsewhere.

\section{The third model}

The field equations resulting from the variation of the Lagrangian with 
respect to the tetrad fields are symmetric (in the $\beta\mu$ indices, 
according to the discussion at the end of section 3)
if we replace the Weitzenb\"ock connection in eq. (\ref{25}) by the 
Christoffel symbols $^0\Gamma^\lambda_{\mu\nu}$. Therefore, in this section
we also assume eqs. (\ref{22}), (\ref{23}), (\ref{24}) and (\ref{25}) of
section 3, as well as the Lagrangian density (\ref{27}), but eq. (\ref{25})
is now constructed out of the Christoffel symbols. This change characterizes
the third model. The field equations of this model, resulting from 
variations of the Lagrangian with respect to $S^\lambda\,_\mu$ and
$e^a\,_\mu$, are given by

\begin{equation}
\nabla_\mu(e\Phi_\lambda\,^{\mu\nu})=0\,,
\label{34}
\end{equation}
where

\begin{equation}
\nabla_\mu(e\Phi_\lambda\,^{\mu\nu})=\partial_\mu(e\Phi_\lambda\,^{\mu\nu})
-e\,^0\Gamma^\sigma_{\mu\lambda}\Phi_\sigma\,^{\mu\nu}\,,
\label{35}
\end{equation}
and

\begin{eqnarray}
&{}& e_{a\nu} e_{b\mu}\partial_\sigma(e\Sigma^{b\nu\sigma})-
e(\Sigma_{\lambda a}\,^\nu T^\lambda\,_{\mu\nu}-
{1\over 4}e_{a\mu}\Sigma^{bcd}T_{bcd}) \nonumber \\
&{}&= -{1\over 4}e\,e_{a\mu}\Phi^{\lambda\rho\nu}{\cal T}_{\lambda\rho\nu}
-{1\over 2}e_a\,^\lambda E_{(\lambda\mu)} \,,
\label{36}
\end{eqnarray}
where

\begin{eqnarray}
E_{\lambda\mu}&=&e(\Phi_\lambda\,^{\rho\nu}{\cal T}_{\mu\rho\nu}-
2\Phi^{\rho\nu}\,_\mu {\cal T}_{\rho\nu\lambda}) \nonumber \\
&-&2e\,g_{\lambda\alpha}g_{\mu\sigma}
(\Phi^{\alpha\rho\beta}S^\nu\,_\beta +
\Phi^{\alpha\nu\beta}S^\rho\,_\beta)\,^0\Gamma^\sigma_{\rho\nu}\nonumber \\
&-&2g_{\lambda\alpha}g_{\mu\sigma}  \biggl[ 
\partial_\nu \biggl( e ( \Phi^{\alpha\nu\beta}S^\sigma\,_\beta
+\Phi^{\alpha\sigma\beta} S^\nu\,_\beta ) \biggr) \nonumber \\
&-&{1\over 2}\partial_\nu \biggl(e ( 
\Phi^{\nu\alpha\beta} S^\sigma\,_\beta+
\Phi^{\nu\sigma\beta} S^\alpha\,_\beta ) \biggl) \biggr]\,, 
\label{37}
\end{eqnarray}
In the last quantity on the right hand side of eq. (\ref{36}), the symbol 
$(\lambda\mu)$ means symmetrization in the indices $\lambda$ and $\mu$. 

As expected, the field equations (\ref{34}) are equivalent to eqs. (\ref{11})
and (\ref{28}). Thus, eq. (\ref{34}) reduces to eqs. (\ref{13}), (\ref{15})
and (\ref{19}) in the flat space-time formulation of the theory, and
ultimately eq. (\ref{34}) describes the propagation of massless spin 2 
fields. Similarly to the model in section 3, 
the left hand side of eq. (\ref{36}) is equivalent to 
Einstein's tensor (except for a factor $1/2$). Equation (\ref{36}) 
reduces to Einstein's equation in vacuum in the limit 
$S^\lambda\,_\mu\rightarrow \delta^\lambda_\mu$.
In spite of being considerably more intricate than eq. (\ref{30}), equation
(\ref{36}) is covariant under local Lorentz transformations, and strictly 
invariant if we convert the free Lorentz index $a$ into a space-time index.
Finally, as asserted at the beginning of this section, 
the field eq. (\ref{36}) is symmetric if the free
Lorentz index is converted into a space-time index.

\section{Discussion}

We have speculated on possible generalizations of the TEGR. The model 
described by eqs. (\ref{6}), (\ref{8}) and (\ref{9}) is no longer
equivalent to the standard general relativity. In addition to the tetrad 
fields, the theory contains the tensor $S^\lambda\,_\mu$ 
that, in the flat space-time formulation, obeys the ordinary wave equation in
vacuum for a spin 2 field. This flat space-time formulation is determined by
eqs. (\ref{6}), (\ref{7}) 
and ${\cal T}^{\lambda}\,_{\mu\nu}=\partial_\mu S^\lambda\,_\nu-
\partial_\nu S^\lambda\,_\mu$, and yields the wave equation
(\ref{19}), which may describe the propagation
of gravitons in a quantum formulation of the theory. This theory is a 
by-product of the more general theory determined by eqs. (\ref{2}), (\ref{6})
and (\ref{7}). 

The formalism presented here allows to make a clear 
distinction between the propagation of gravitational waves and of spin 2 
fields, an issue that is not properly and satisfactorily 
addressed in the ordinary formulation of
general relativity. In the framework of the coupled field equations
(\ref{8}) and (\ref{9}), the source of the field $S^{\lambda\mu}$ is the
dynamical geometry of the space-time, represented by time dependent
tetrad fields $e^a\,_\mu$. 

In the realm of the theory defined by 
eqs. (\ref{6}), (\ref{8}) and (\ref{9}),
if we take the limit $S^\lambda\,_\mu\rightarrow \delta^\lambda_\mu$, we must
necessarily make $e^a\,_\mu\rightarrow\delta^a_\mu$ simultaneously,
otherwise the tetrad fields would be over-determined. This difficulty (or in
fact, this feature) does not take place in the models addressed in 
sections 3 and 4.

In the framework of eqs. (\ref{27}), (\ref{28}) and (\ref{30}) (second 
model), or (\ref{34}) and (\ref{36}) (third model) , by taking the limit 
$S^\lambda\,_\mu\rightarrow\delta^\lambda_\mu$ we arrive at the standard
form of the vacuum Einstein's equations in the TEGR. In contrast, 
the model established
in section 2 may be thought as a significant change of paradigm regarding the
ordinary formulation of general relativity, since the establishment of the 
flat space-time is more subtle and unconventional. Of course, as a possible 
physical realization, we may have a nearly
flat space-time described by the tetrad fields, with non-vanishing 
background weak oscillating fields $S^\lambda\,_\mu$.

An issue that certainly deserves a careful investigation is the interaction
of both $e^a\,_\mu$ and $S^\lambda\,_\mu$ with matter fields. The analysis 
of this issue could ensure or not the viability of the models presented in
sections 2, 3 and 4.

Quantization in the teleparallel geometry has already been carried out in
references \cite{Ulhoa1,Ulhoa2}. It is possible that the techniques
developed in the latter references could by applied to $S^{\lambda\mu}$ in
a simplified framework where $e_{a\mu}$ represents a weak gravitational field.

Finally, an interesting issue to be addressed is the establishment 
of a possible dynamics for the anti-symmetric part of the tensor 
$S^{\lambda\mu}$, since this quantity is not fixed by the field equations
(\ref{13}).

\end{document}